\documentstyle[12pt]{article}

\newcommand{\1}{{\sf 1 \!\! 1}}
\newcommand{\da}{^\dagger}
\makeatletter
\@addtoreset{equation}{section}
\makeatother

\title{Quantum Link Models: \\ A Discrete Approach to Gauge Theories
\footnote{This work is supported in part by funds provided by the U.S.
Department of Energy (D.O.E.) under cooperative research agreement
DE-FC02-94ER40818.}}
\author{S. Chandrasekharan and U.-J. Wiese \\ \\
Center for Theoretical Physics, \\
Laboratory for Nuclear Science, and Department of Physics \\
Massachusetts Institute of Technology (MIT) \\
Cambridge, Massachusetts 02139, U.S.A. \\ \\
MIT Preprint, CTP 2573 \\ \\}

\begin{document} 
\maketitle
\begin{abstract} \normalsize
 
We construct lattice gauge theories in which the elements of the link matrices 
are represented by non-commuting operators acting in a Hilbert space. These 
quantum link models are related to ordinary lattice gauge theories in the same 
way as quantum spin models are related to ordinary classical spin systems. 
Here $U(1)$ and $SU(2)$ quantum link models are constructed explicitly. As 
Hamiltonian theories quantum link models are nonrelativistic gauge theories 
with potential applications in condensed matter physics. When formulated with 
a fifth Euclidean dimension, universality arguments suggest that dimensional 
reduction to four dimensions occurs. Hence, quantum link models are also 
reformulations of ordinary quantum field theories and are  
applicable to particle physics, for example to QCD. The configuration space of 
quantum link models is discrete and hence their numerical treatment 
should be simpler than that of ordinary lattice gauge theories with a 
continuous configuration space.

\end{abstract}
 
\maketitle
 
\newpage

\section{Introduction}

Gauge theories provide the fundamental structure that is used to describe
the interactions of elementary particles. For example, the dynamics of the 
Standard model is formulated as an $SU(3)\otimes SU(2)\otimes U(1)$ gauge 
theory. Quantum chromodynamics (QCD) --- the $SU(3)$ gauge theory of quarks and
gluons --- is strongly coupled, and hence requires a nonperturbative 
formulation. At present the only formulation of this kind is provided by the
lattice regularization, in which a gauge theory formally resembles a classical 
statistical mechanics system. The gauge degrees of freedom are then represented
by parallel transporters, which are matrices taking values in the gauge group,
and which are naturally associated with the links connecting neighboring 
lattice sites. Gauge symmetries also arise in condensed matter systems. For
example, in superconductors the spontaneous break down of the $U(1)$ gauge 
symmetry of quantum electrodynamics (QED) plays a central role.
Furthermore, effective gauge symmetries may be dynamically generated, even 
though they are not present at the fundamental level of the Standard model. 
For example, in nonrelativistic quantum Hall fluids an $SU(2)$ gauge symmetry 
results from the coupling of orbital and spin angular momenta \cite{Fro93}. In
the context of condensed matter physics a physical crystal lattice often 
serves as a regulator. In these cases a nonrelativistic lattice gauge 
Hamiltonian can be used to describe the system. Lattice gauge symmetries may 
arise even at macroscopic scales, as for example in man-made arrays of 
Josephson junctions \cite{Eck89}.

Both in elementary particle and in condensed matter physics gauge theories are 
usually formulated using path integrals. Then the gauge degrees of freedom are 
described by classical fields, and the theory is analogous to a system of
classical statistical mechanics. Here we formulate lattice gauge theories such
that the classical statistical mechanics problem is converted into
one of quantum statistical mechanics. In other words, the classical Hamilton
function (or classical Euclidean action in the context of field theory) is
replaced by a Hamilton operator. As a consequence, the elements of the link 
matrices that are ordinary c-numbers in the standard formulation of lattice 
gauge theory now turn into non-commuting operators acting in a Hilbert space.
In the context of spin models this kind of quantization is well established 
when one goes from a classical to a quantum spin system. For example, the 
classical $O(3)$ spin model with 3-component classical unit vectors on each 
lattice site is replaced by the quantum Heisenberg model, in which each spin is
represented by a vector of Pauli matrices \cite{Hei28}. In the same sense, 
quantum link models are quantized versions of ordinary classical lattice 
gauge theories. Such models were first constructed by Horn \cite{Hor81} in
1981. In 1990 they were rediscovered under the name of gauge magnets and
investigated in more detail by Orland and Rohrlich \cite{Orl90,Orl92}. 
\footnote{We thank P. Orland for drawing our attention to 
refs.\cite{Hor81,Orl90,Orl92}, which we were not aware of in the preprint 
version of this paper.} Having rediscovered these models another time, we go 
beyond the earlier work by showing how quantum link models are related to 
ordinary quantum field theories via dimensional reduction, and how they may be 
used to improve numerical simulations of QCD.

The Hamilton operator of a quantum link model resembles the Hamilton
function of the corresponding classical system. By construction such a
system is nonrelativistic. At present we can only speculate about potential
applications of quantum link models to real condensed matter systems. Since a 
variety of models becomes available, it is of theoretical
interest to study their properties, and perhaps relate them to some 
experimental phenomenon, like, for example, high $T_c$ superconductivity. 
In any case, quantum link models possess a rich mathematical structure, and
they are as general as ordinary lattice gauge theories. Hence, one may expect
that Nature has made use of them at some level.

Here we concentrate on the application of quantum link models to elementary 
particle
physics. In this context, the Hamilton operator of the quantum link model
replaces the Euclidean action of the corresponding ordinary lattice gauge
theory. By construction the symmetries of the quantum link model are the same
as the ones of the corresponding ordinary lattice gauge theory. Therefore, if
one can take the continuum limit of a quantum link model, universality 
arguments suggest that it corresponds to the same continuum field theory
as its classical counterpart. In this context the Hamilton operator of a 
quantum link model describes the evolution of the system in a fifth continuous
Euclidean dimension that is distinct from Euclidean time, which is part of the 
lattice. The extent of the extra dimension resembles the inverse coupling 
constant of the corresponding 4-d lattice gauge theory. Hence, varying the 
extent of the 
extra dimension allows one to approach the continuum limit. In this limit 
dimensional reduction to four dimensions occurs. This suggests that quantum 
link models are also a reformulation of ordinary gauge theories, and hence are
applicable to theories like QCD. 

The configuration space of a quantum link model is discrete. Therefore 
numerical approaches to quantum link models should be simpler than in ordinary
lattice gauge theories. In particular, it is easier to store large lattices, 
and it should require less computer time to generate configurations. Also it is
conceivable that efficient numerical techniques, for example cluster 
algorithms, which 
do not work for ordinary lattice gauge theories, become available for quantum 
link models. 

The paper is organized as follows. In section 2 we contrast classical with
quantum spin systems, and we discuss in which sense the 2-d antiferromagnetic
quantum Heisenberg model can be viewed as a discrete realization of the 
ordinary classical 2-d $O(3)$ model. In section 3 we construct the simplest
quantum link model with a $U(1)$ gauge symmetry. Section 4 contains the
construction of an $SU(2)$ quantum link model. In section 5 we formulate
quantum link models with a fifth Euclidean dimension and discuss their
dimensional reduction to ordinary 4-d quantum field theories. Finally, section 
6 contains our conclusions.

\section{Classical versus Quantum Spin Models}

In this section we discuss quantum models in the well established context of
spin systems, and we discuss how they are related to their classical 
counterparts. Later we will generalize these structures to gauge theories.
Let us consider standard classical $O(N)$ symmetric spin systems on a 
$d$-dimensional cubic lattice with a classical $N$-component unit vector 
$\vec s_x$ attached to each lattice point $x$. We discuss this problem in the
language of lattice field theory with the classical action (or in the language
of classical statistical mechanics the classical Hamilton function) of this 
model given by
\begin{equation}
S[\vec s] = - \sum_{x,\mu} \vec s_x \cdot \vec s_{x+\hat\mu},
\end{equation}
where $\hat\mu$ represents the unit vector in the $\mu$-direction. One is
interested in the partition function
\begin{equation}
Z = \int {\cal D}\vec s \ \exp(- \frac{1}{g} S[\vec s]),
\end{equation}
which is a path integral over all classical spin field configurations 
$[\vec s]$. 
Here $g$ is the coupling constant (or equivalently the temperature in classical
statistical mechanics language). The $N=2$ case is the $XY$-model, and $N=3$ 
corresponds to the classical $O(3)$ model. The 2-d $XY$-model has a 
Kosterlitz-Thouless transition at a finite value of $g$, separating a massless
spin-wave phase at small coupling from a massive strong coupling phase with a 
vortex condensate. One can take a continuum limit of this lattice model 
anywhere in the spin-wave phase. The resulting continuum field theory describes
a free massless boson. The 2-d $O(3)$-model, on the other hand, has only one 
phase with a nonperturbatively generated mass gap. The mass gap
vanishes exponentially as $g$ goes to zero. Hence, the corresponding continuum
field theory is asymptotically free.

Heisenberg constructed a quantum version of the $O(3)$-model by replacing the 
classical spins $\vec s_x$ by quantum spin operators $\vec S_x$. The classical 
action (or Hamilton function) then turns into the Hamilton operator
\begin{equation}
H = J \sum_{x,\mu} \vec S_x \cdot \vec S_{x+\hat\mu}.
\end{equation}
For $J < 0$ we have a ferromagnet, while $J > 0$ corresponds to an 
antiferromagnet. It is important to realize that the components of the spin 
vectors obey the standard commutation relations
\begin{equation}
[S_x^i,S_y^j] = i \delta_{xy} \epsilon_{ijk} S_x^k.
\end{equation}
Of course, spin operators located at different lattice points commute with each
other. Consequently, the above Hamilton operator commutes with the total spin
\begin{equation}
\vec S = \sum_x \vec S_x,
\end{equation}
i.e. $[\vec S,H] = 0$, and hence, like the classical model, the quantum spin 
model has a global $O(3)$ symmetry. Now one is interested in the quantum 
partition function
\begin{equation}
Z = \mbox{Tr} \exp(- \beta H).
\end{equation}
The trace is taken in the Hilbert space, which is a direct product of the
Hilbert spaces of individual spins. 

Quantum spin models can be realized with various representations of the spin. 
The corresponding theories may behave quite differently. For example, Haldane 
has conjectured that 1-d antiferromagnetic $O(3)$ quantum spin chains with 
integer spins have a mass gap, while those with half-integer spins are gapless 
\cite{Hal83}. In fact, the spin 1/2 antiferromagnetic Heisenberg chain has been
solved by the Bethe ansatz, and indeed turns out to have no mass gap 
\cite{Bet31}. The same has been shown analytically for all half-integer spins
\cite{Lie61}. On the other hand, there is strong numerical evidence for a 
mass gap in spin 1 and spin 2 systems \cite{Bot84}. In the classical limit of 
large spin $S$ the mass gap vanishes as $m \propto \exp(- \pi S)$ thus 
approaching a continuum limit. It is interesting that 1-d antiferromagnetic 
quantum spin chains can be mapped to the 2-d classical $O(3)$-model with a 
$\theta$-vacuum term. Integer spins correspond to $\theta = 0$. Then the 2-d 
$O(3)$-model 
indeed has a mass gap that vanishes as $m \propto \exp(- 2 \pi/g)$ in the 
low temperature limit. Hence, for large $S$ one may identify the spin of the 
1-d quantum model, $S = 2/g$, with the inverse coupling of 
the 2-d classical model. Half-integer spins correspond to $\theta=\pi$ and it 
turns out that the mass gap of the $O(3)$-model then disappears \cite{Bie95}. 
In that case the 1-d quantum model corresponds to a 2-d conformal field theory 
--- the $k=1$ Wess-Zumino-Novikov-Witten model \cite{Nov81} --- as was first
argued by Affleck \cite{Aff88}.

Also the 2-d antiferromagnetic spin 1/2 quantum Heisenberg model has very
interesting properties. First of all, it describes the precursor insulators
of high $T_c$ superconductors --- materials like $\mbox{La}_2\mbox{CuO}_4$ ---
whose ground states are N\'eel ordered with a spontaneously generated staggered
magnetization. Indeed, there has been early numerical evidence that the ground
state of the 2-d antiferromagnetic spin 1/2 quantum Heisenberg model shows 
long range order \cite{Bar91}. This has been confirmed by a very precise
numerical study using a loop cluster algorithm \cite{Wie94}. Recently, the
loop cluster algorithm has been reformulated to work in the Euclidean time
continuum, allowing high-precision studies of the extreme low temperature
limit \cite{Bea96}. 

Formulating the 2-d quantum model as a path integral in
Euclidean time results in a 3-d $O(3)$-symmetric classical model. At zero 
temperature of the quantum system we are in the infinite volume limit of the
corresponding 3-d classical model. The N\'eel order of the ground state of the 
2-d quantum system implies that the corresponding 3-d classical system is in 
the broken phase with massless Goldstone bosons --- in this case 
two antiferromagnetic magnons (or spin-waves). One can use chiral perturbation
theory to describe the dynamics of the Goldstone bosons at low energies
\cite{Leu90}. To lowest order the effective action then takes the form
\begin{equation}
\label{spinaction}
S[\vec s] = \int_0^\beta dt \int d^2x \ \frac{\rho_s}{2}
[\partial_\mu \vec s \cdot \partial_\mu \vec s + 
\frac{1}{c^2} \partial_t \vec s \cdot \partial_t \vec s].
\end{equation}
Here $c$ and $\rho_s$ are the spin-wave velocity and the spin stiffness.
The 2-d quantum system at finite temperature corresponds to the 3-d classical 
model with finite Euclidean time extent
$\beta$. For massless particles --- in our case the Goldstone bosons ---
the finite temperature system appears dimensionally reduced to two dimensions,
because the finite Euclidean time extent is then negligible compared to the
correlation length. However, the Mermin-Wagner-Coleman theorem prevents the
existence of interacting massless Goldstone bosons in two dimensions 
\cite{Mer66}. Indeed, the 2-d $O(3)$ model has a nonperturbatively generated
mass gap. Hasenfratz and Niedermayer used a block spin
renormalization group transformation to map the 3-d $O(3)$-model with finite
Euclidean time extent $\beta$ to a 2-d lattice $O(3)$-model \cite{Has91}.
They averaged the 3-d field over space-time volumes of size $\beta$ in
the Euclidean time direction and $\beta c$ in the two space directions. 
Due to the large correlation length the field
is essentially constant over these blocks. The averaged field naturally lives
at the block centers, which form a 2-d lattice of spacing $\beta c$ (which is
different from the lattice spacing of the underlying quantum antiferromagnet).
Hence the effective action of the averaged field defines a 2-d classical
lattice $O(3)$-model. Using chiral perturbation theory, Hasenfratz and
Niedermayer expressed its coupling constant as
\begin{equation}
1/g = \beta \rho_s - \frac{3}{16 \pi^2 \beta \rho_s} + 
{\cal O}(1/\beta^2 \rho_s^2).
\end{equation}
Using the 3-loop $\beta$-function of the 2-d $O(3)$-model together with its
exact mass gap \cite{Has90}, they also extended an earlier result of 
Chakravarty, Halperin and Nelson \cite{Cha89} for the inverse correlation 
length of the quantum antiferromagnet to
\begin{equation}
m = \frac{16 \pi \rho_s}{e c} \exp(- 2 \pi \beta \rho_s)
[1 + \frac{1}{4 \pi \beta \rho_s} + {\cal O}(1/\beta^2 \rho_s^2)].
\end{equation}
Here $e$ is the base of the natural logarithm. The above equation resembles
the asymptotic scaling behavior of the 2-d classical $O(3)$-model. Hence, one
can view the 2-d antiferromagnetic quantum $O(3)$-model in the zero 
temperature
limit as a reformulation of the 2-d classical model. It is remarkable
that this formulation is entirely discrete, even though the classical model is
usually formulated with a continuous configuration space. Further, the quantum
model can be treated with very efficient loop cluster algorithm techniques
\cite{Eve93,Wie94}. Defining the path integral for discrete quantum systems 
does not require discretization of Euclidean time. This observation has 
recently led to a very efficient loop cluster algorithm operating directly in
the Euclidean time continuum \cite{Bea96}. Of course, for the classical 
$O(3)$-model the Wolff cluster algorithm is also available \cite{Wol89}.

A quantum spin model with $O(2)$ symmetry 
has also been constructed and is known as the quantum $XY$ model. Its Hamilton 
operator is given by
\begin{equation}
H = J \sum_{x,\mu} [S_x^1 S_{x+\hat\mu}^1 + S_x^2 S_{x+\hat\mu}^2],
\end{equation}
in analogy with the corresponding classical model. In this case $H$ only 
commutes with the third component of the total spin, i.e. $[S^3,H] = 0$.
The phenomenon of dimensional reduction also occurs in the quantum 
$XY$-model. There is numerical evidence for a Kosterlitz-Thouless transition
at finite temperature \cite{Din90} just like in the classical $XY$-model. 
In the low temperature phase the theory describes free massless spin-waves.
Hence, again the finite Euclidean time extent of the quantum model is 
negligible compared to the correlation length. In contrast to the $O(3)$-model
no mass gap is generated, and the Mermin-Wagner-Coleman theorem is evaded,
because the massless particles do not interact in this case. 

In the following we construct 4-d quantum link models with a gauge symmetry. 
When formulated as 5-d classical gauge theories with a finite extent 
in the fifth dimension, these models can be viewed as reformulations of 
ordinary 4-d gauge theories. Again, the configuration space of the
quantum models is entirely discrete, and one may hope that cluster algorithms
become available, even though they don't work in the standard formulation.

Finally, let us comment on quantum ferromagnets. These systems have a highly
degenerate ground state (even in a finite volume) and a conserved order
parameter. As a consequence, the dispersion relation of the corresponding
Goldstone bosons is nonrelativistic, and the arguments from above do not
apply. 

\section{The $U(1)$ Quantum Link Model}

Let us discuss the simplest quantum link model --- quantum pure $U(1)$ gauge
theory. The corresponding classical model has a $U(1)$ parallel transporter
\begin{equation}
u_{x,\mu} = \exp(i\varphi_{x,\mu}) = 
\cos\varphi_{x,\mu} + i \sin\varphi_{x,\mu},
\end{equation}
associated with each link $x,\mu$. The classical lattice action is given by
\begin{equation}
S[u] = - \frac{1}{2} \sum_{x,\mu > \nu} [u_{x,\mu} u_{x+\hat\mu,\nu} 
u_{x+\hat\nu,\mu}\da u_{x,\nu}\da + 
u_{x,\nu} u_{x+\hat\nu,\mu} u_{x+\hat\mu,\nu}\da u_{x,\mu}\da],
\end{equation}
where a dagger denotes complex conjugation. By construction, the action is
invariant under $U(1)$ gauge transformations
\begin{equation}
u'_{x,\mu} = \exp(i \alpha_x) u_{x,\mu} \exp(- i \alpha_{x+\hat\mu}).
\end{equation}
One is interested in the partition function as a path integral over
classical link configurations
\begin{equation}
Z = \int {\cal D}u \ \exp(- \frac{1}{g^2} S[u]).
\end{equation}
Here $g$ is the gauge coupling. Formally we can think about the system as one 
of classical statistical mechanics. Then the action $S[u]$ plays
the role of the classical Hamilton function, and $g^2$ plays the role of the
temperature. The 4-d $U(1)$ lattice gauge theory has a phase transition
separating a massless Coulomb phase at weak coupling from a massive confined 
phase with condensed monopoles at large $g$. The continuum limit of this
model can be taken anywhere in the Coulomb phase, resulting in a theory of
free massless photons. This is in close analogy to the 2-d classical 
$XY$-model.

Let us now construct the quantum counterpart of the standard $U(1)$ gauge
theory. We want to replace the classical action by a quantum Hamilton operator
\begin{equation}
H = \frac{J}{2} \sum_{x,\mu > \nu} [U_{x,\mu} U_{x+\hat\mu,\nu} 
U_{x+\hat\nu,\mu}\da U_{x,\nu}\da + 
U_{x,\nu} U_{x+\hat\nu,\mu} U_{x+\hat\mu,\nu}\da U_{x,\mu}\da],
\end{equation}
where $U_{x,\mu}$ now is an operator acting in a Hilbert space --- not just
a c-number. In quantum mechanics there is no Hermitean operator that replaces
a classical angle $\varphi_{x,\mu}$. Instead, one should work with 
$\cos\varphi_{x,\mu}$ and $\sin\varphi_{x,\mu}$. Hence, we write
\begin{equation}
U_{x,\mu} = C_{x,\mu} + i S_{x,\mu},
\end{equation}
where $C_{x,\mu}$ and $S_{x,\mu}$ are Hermitean operators yet to be determined.
Consequently, we write
\begin{equation}
U_{x,\mu}\da = C_{x,\mu} - i S_{x,\mu}.
\end{equation}
Here the dagger represents Hermitean conjugation in the Hilbert space.
The gauge symmetry of the quantum link model requires that the above Hamilton 
operator commutes with the generators $G_x$ of infinitesimal gauge 
transformations at each lattice site $x$. This is satisfied by construction if 
the quantum link operator transforms as
\begin{equation}
U'_{x,\mu} = \exp(- i \alpha_x G_x) U_{x,\mu} \exp(i \alpha_x G_x) =
\exp(i \alpha_x) U_{x,\mu},
\end{equation}
under gauge transformations from the left, and as
\begin{equation}
U'_{x,\mu} = \exp(- i \alpha_{x+\hat\mu} G_{x+\hat\mu}) U_{x,\mu} 
\exp(i \alpha_{x+\hat\mu} G_{x+\hat\mu}) = 
U_{x,\mu} \exp(- i \alpha_{x+\hat\mu}),
\end{equation}
under gauge transformations from the right. The unitary operator that 
represents a general gauge transformation then is
$\prod_x \exp(i \alpha_x G_x)$ such that
\begin{equation}
U'_{x,\mu} = \prod_y \exp(- i \alpha_y G_y) U_{x,\mu} 
\prod_z \exp(i \alpha_z G_z) = \exp(i \alpha_x) U_{x,\mu}
\exp(- i \alpha_{x+\hat\mu}).
\end{equation}
The above structure implies the following commutation relations
\begin{eqnarray}
&&[G_x,C_{y,\mu}] = i (\delta_{x,y+\hat\mu} - \delta_{x,y}) S_{y,\mu},
\nonumber \\
&&[G_x,S_{y,\mu}] = i (\delta_{x,y} - \delta_{x,y+\hat\mu}) C_{y,\mu},
\end{eqnarray}
and hence
\begin{eqnarray}
&&[G_x,U_{y,\mu}] = (\delta_{x,y+\hat\mu} - \delta_{x,y}) U_{y,\mu},
\nonumber \\
&&[G_x,U_{y,\mu}\da] = (\delta_{x,y} - \delta_{x,y+\hat\mu}) U_{y,\mu}\da.
\end{eqnarray}
It is straightforward to show that these relations are satisfied
when one identifies
\begin{equation}
C_{x,\mu} = S_{x,\mu}^1, \ S_{x,\mu} = S_{x,\mu}^2,
\end{equation}
and
\begin{equation}
G_x = \sum_\mu (S_{x-\hat\mu,\mu}^3 - S_{x,\mu}^3),
\end{equation}
where $\vec S_{x,\mu}$ obeys angular momentum commutation relations, i.e.
\begin{equation}
[S_{x,\mu}^i,S_{y,\nu}^j] = i \delta_{x,y} \delta_{\mu\nu} \epsilon_{ijk}
S_{x,\mu}^k.
\end{equation}
We can now identify
\begin{eqnarray}
&&U_{x,\mu} = C_{x,\mu} + i S_{x,\mu} = S^1_{x,\mu} + i S^2_{x,\mu} =
S^+_{x,\mu}, \nonumber \\
&&U_{x,\mu}\da = C_{x,\mu} - i S_{x,\mu} = S^1_{x,\mu} - i S^2_{x,\mu} =
S^-_{x,\mu},
\end{eqnarray}
i.e. a link variable is represented by a raising operator $S^+_{x,\mu}$, and 
its inverse by the lowering operator $S^-_{x,\mu}$. Like for quantum spin 
systems, the above commutation relations can be realized
with any representation of $SU(2)$. In the simplest case one can use Pauli
matrices on each link. Then the Hilbert space of the model is the direct
product of 2-dimensional link Hilbert spaces. 

By construction the above Hamilton operator is invariant under gauge
transformations, i.e. $[G_x,H] = 0$ for all $x$. Also the generators of gauge
transformations commute, i.e. $[G_x,G_y] = 0$. Hence, the
eigenstates of $H$ can be characterized by the eigenvalues of all $G_x$, i.e.
there is a conserved quantity at each lattice site. In gauge theories Gauss'
law restricts the physical Hilbert space to gauge invariant states 
$|\Psi\rangle$. In our formulation this means
\begin{equation}
G_x |\Psi\rangle = 0.
\end{equation}

Let us contrast our construction with the Hamiltonian formulation of ordinary
lattice gauge theories. There one often chooses an electric flux basis of the
physical Hilbert space. In an ordinary $U(1)$ lattice gauge theory the electric
flux is quantized in integer units. Gauss' law requires that the fluxes 
associated with links emanating from the same lattice point add up to zero.
The electric part of the Hamiltonian is diagonal in the electric flux basis.
The magnetic part associated with spatial plaquettes, on the other hand, 
induces a shift of the electric flux on all links of the plaquette. This is
in close analogy to the $U(1)$ quantum link model. In particular, one can 
identify the eigenvalues of $S^3_{x,\mu}$ with 
electric fluxes associated with the links. Again, the Hamiltonian induces
shifts in the electric fluxes around a plaquette. However, in this case the
fluxes, being the eigenvalues of $S^3_{x,\mu}$, are restricted to a finite set
(for example to $\pm 1/2$ when one chooses the fundamental representation of 
$SU(2)$). Thus, in contrast to ordinary lattice gauge theories the Hilbert 
space of a quantum link model is finite (on a finite lattice).

At this point we have constructed what we call the $U(1)$ quantum link model.
As for ordinary lattice gauge models, solving the theory is a complicated 
problem. In two dimensions this has been discussed in ref.\cite{Orl92}.
Before we discuss the dynamics let us construct an example of a
nonabelian quantum link model.

\section{The $SU(2)$ Quantum Link Model}

Let us first recall standard $SU(2)$ lattice gauge theory. In that case there 
is an $SU(2)$ matrix
\begin{equation}
\label{classicallink}
u_{x,\mu} = u^0_{x,\mu} + i \vec u_{x,\mu} \cdot \vec \sigma,
\end{equation}
associated with each link. Here $\vec \sigma$ is a vector of Pauli matrices.
Further, $u^0_{x,\mu}$ and $\vec u_{x,\mu}$ are real and satisfy the constraint
$u^0_{x,\mu} u^0_{x,\mu} + \vec u_{x,\mu} \cdot \vec u_{x,\mu} = 1$. The
action is given by
\begin{equation}
S[u] = - \sum_{x,\mu > \nu} \mbox{Tr} [u_{x,\mu} u_{x+\hat\mu,\nu} 
u_{x+\hat\nu,\mu}\da u_{x,\nu}\da + u_{x,\nu} u_{x+\hat\nu,\mu} 
u_{x+\hat\mu,\nu}\da u_{x,\mu}\da],
\end{equation}
where the dagger denotes Hermitean conjugation. The action is invariant under
$SU(2)$ gauge transformations
\begin{equation}
u'_{x,\mu} = \exp(i \vec \alpha_x \cdot \vec \sigma) u_{x,\mu}
\exp(- i \vec \alpha_{x+\hat\mu} \cdot \vec \sigma).
\end{equation}
Again, the path integral is given by
\begin{equation}
Z = \int {\cal D}u \exp(- \frac{1}{g^2} S[u]),
\end{equation}
where $g$ is the nonabelian gauge coupling. In contrast to $U(1)$ gauge theory,
$SU(2)$ gauge theory in four dimensions has only one phase, in which the gluons
are confined. This is analogous to the 2-d $O(3)$-model.

As in the $U(1)$ case we can construct a quantum version of the $SU(2)$ model
by replacing the classical action by a Hamilton operator
\begin{equation}
H = J \sum_{x,\mu > \nu} \mbox{Tr} [U_{x,\mu} U_{x+\hat\mu,\nu} 
U_{x+\hat\nu,\mu}\da U_{x,\nu}\da + U_{x,\nu} U_{x+\hat\nu,\mu} 
U_{x+\hat\mu,\nu}\da U_{x,\mu}\da].
\end{equation}
Here the elements of the $2 \times 2$ link matrices $U_{x,\mu}$ are operators 
acting in a Hilbert space. Naturally, the dagger now represents Hermitean 
conjugation in both the Hilbert space and the $2 \times 2$ matrix space. In 
analogy to the classical expression of eq.(\ref{classicallink}) we write
\begin{equation}
U_{x,\mu} = U^0_{x,\mu} + i \vec U_{x,\mu} \cdot \vec \sigma,
\end{equation}
where $U^0_{x,\mu}$ and $\vec U_{x,\mu}$ are Hermitean operators. We also have
\begin{equation}
U_{x,\mu}\da = U^0_{x,\mu} - i \vec U_{x,\mu} \cdot \vec \sigma.
\end{equation}
In analogy with the $U(1)$ case, gauge covariance requires
\begin{equation}
U'_{x,\mu} = \prod_y \exp(- i \vec \alpha_y \cdot \vec G_y) U_{x,\mu} 
\prod_z \exp(i \vec \alpha_z \cdot \vec G_z) = 
\exp(i \vec \alpha_x \cdot \vec \sigma) U_{x,\mu}
\exp(- i \vec \alpha_{x+\hat\mu} \cdot \vec \sigma).
\end{equation} 
This implies the following commutation relations
\begin{eqnarray}
\label{GUcommutator}
&&[\vec G_x,U_{y,\mu}] = \delta_{x,y+\hat\mu} U_{y,\mu} \vec \sigma -
\delta_{x,y} \vec \sigma U_{y,\mu},
\nonumber \\
&&[\vec G_x,U_{y,\mu}\da] = \delta_{x,y} U_{y,\mu}\da \vec \sigma -
\delta_{x,y+\hat\mu} \vec \sigma U_{y,\mu}\da 
\end{eqnarray}
One way to find representations that satisfy these relations is to define
\begin{equation}
\vec G_x = \sum_\mu (\vec R_{x-\hat\mu,\mu} + \vec L_{x,\mu}).
\end{equation}
Here $\vec R_{x,\mu}$ and $\vec L_{x,\mu}$ are generators of right and left 
gauge transformations of the link variable $U_{x,\mu}$. As such, they obey
the following commutation relations
\begin{eqnarray}
&&[R_{x,\mu}^i,R^j_{y,\nu}] = 2 i \delta_{x,y} \delta_{\mu\nu} 
\epsilon_{ijk} R_{x,\mu}^k, \nonumber \\
&&[L_{x,\mu}^i,L^j_{y,\nu}] = 2 i \delta_{x,y} \delta_{\mu\nu} 
\epsilon_{ijk} L_{x,\mu}^k, \nonumber \\
&&[R_{x,\mu}^i,L^j_{y,\nu}] = 0,
\end{eqnarray}
i.e. $\vec R_{x,\mu}$ and $\vec L_{x,\mu}$ generate an 
$SU(2)_R \otimes SU(2)_L$ algebra on each link.
The commutation relations of eq.(\ref{GUcommutator}) imply
\begin{eqnarray}
&&[\vec R_{x,\mu},U_{y,\nu}] = \delta_{x,y} \delta_{\mu\nu} U_{x,\mu} 
\vec \sigma, \nonumber \\
&&[\vec L_{x,\mu},U_{y,\nu}] = - \delta_{x,y} \delta_{\mu\nu} 
\vec \sigma U_{x,\mu}.
\end{eqnarray}
For each link the above relations can be realized by using the generators
of an $SO(5)$ algebra, with the $SU(2)_L \otimes SU(2)_R$ algebra embedded in 
it. In the spinorial representation, for example, the ten generators of 
$SO(5)$ can be written as
\begin{eqnarray}
&&\vec R = \left( \begin{array}{cc} \vec \tau & 0 \\ 0 & 0 \end{array} \right),
\ \vec L = \left( \begin{array}{cc} 0 & 0 \\ 0 & \vec \tau \end{array} \right),
\nonumber \\
&&U^0 = \left( \begin{array}{cc} 0 & \1 \\ \1 & 0 \end{array} \right), \
\vec U = \left( \begin{array}{cc} 0 & - i \vec \tau \\ i \vec \tau & 0 
\end{array} \right).
\end{eqnarray}
Here $\1$ is a $2 \times 2$ unit matrix, and $\vec \tau$ is a vector of
Pauli matrices (not to be confused with $\vec \sigma$, which acts in a 
different space). Note that $(U^0,\vec U)$ resembles a four-vector of Euclidean
Dirac matrices. The commutators of the components of the link matrices then
take the form
\begin{eqnarray}
&&[U_{x,\mu}^0,U_{y,\nu}^0] = 0, \nonumber \\
&&[U_{x,\mu}^0,U_{y,\nu}^i] = 2 i \delta_{x,y} \delta_{\mu\nu} 
(R_{x,\mu}^i - L_{x,\mu}^i), \nonumber \\
&&[U_{x,\mu}^i,U_{y,\nu}^j] = 2 i \delta_{x,y} \delta_{\mu\nu} 
\epsilon_{ijk} (R_{x,\mu}^k + L_{x,\mu}^k).
\end{eqnarray}
The above commutation relations can be realized with any representation of 
$SO(5)$. The simplest choice is the spinorial representation that was used
above. Then the Hilbert space of the model is the direct product of 
4-dimensional link Hilbert spaces. By construction we have $[\vec G_x,H] = 0$.
To impose the Gauss law in the $SU(2)$ case one requires
\begin{equation}
\vec G_x |\Psi \rangle = 0,
\end{equation}
for all physical states $|\Psi \rangle$. 

\section{Reduction from five to four Dimensions}

As we have seen, a 2-d $O(3)$ quantum spin model can be viewed as a discrete 
reformulation of a 2-d classical $O(3)$ spin model, because
dimensional reduction from three to two dimensions occurs. We also know that 
the 2-d quantum $O(3)$-model at finite temperature corresponds to a 3-d 
classical $O(3)$-model with finite extent $\beta$ in the Euclidean time 
direction. Due to the N\'eel order of the underlying 2-d quantum 
antiferromagnet the theory is in the broken phase with massless Goldstone 
bosons. In the leading order of chiral perturbation theory the effective action
of the 3-d model takes the form 
\begin{equation}
S[\vec s] = \int_0^\beta dt \int d^2x \ \frac{\rho_s}{2}
[\partial_\mu \vec s \cdot \partial_\mu \vec s + 
\frac{1}{c^2} \partial_t \vec s \cdot \partial_t \vec s].
\end{equation}
In fact, the extent of the third dimension --- the inverse temperature $\beta$ 
of the quantum antiferromagnet --- resembles the inverse coupling of the 
induced 2-d classical model, i.e. at large $\beta$
\begin{equation}
\label{O3coupling}
1/g = \beta \rho_s.
\end{equation}
Thus, the continuum limit of the asymptotically free 2-d classical
$O(3)$-model at $g \rightarrow 0$ corresponds to the zero temperature limit
of the 2-d quantum antiferromagnet, which also corresponds to the infinite 
volume limit
($\beta \rightarrow \infty$) of the 3-d classical model. When $\beta$ is 
finite, the theory is effectively two dimensional and the Mermin-Wagner-Coleman
theorem implies that the Goldstone bosons then pick up a mass
\begin{equation}
m \propto \exp(- 2 \pi \beta \rho_s).
\end{equation}
This is consistent with the asymptotic freedom of the 2-d classical 
$O(3)$-model. The correlation length $1/m$ is exponentially large compared to 
the Euclidean time extent $\beta$, and hence, in contrast to naive 
expectations, dimensional reduction occurs at large $\beta$.

The above observations imply an interesting relation between $O(3)$-models in
two and three dimensions. In the infinite volume ($\beta = \infty$) the
3-d model describes the physics of massless Goldstone bosons. The corresponding
fixed point of the renormalization group resembles a conformal field theory.
Once we make $\beta$ finite (which implies $g > 0$) we explore a relevant
direction in the vicinity of this fixed point. Approaching the fixed point
along the relevant direction yields the 2-d $O(3)$-model. In the following we 
argue that nonabelian gauge theories in four and five dimensions are related
in a similar way.

Nonabelian lattice gauge theories in five dimensions have a confinement phase
at strong coupling, which is separated from a massless weak coupling 
phase \cite{Cre79}. When a gauge theory is dimensionally reduced, usually the
Polyakov loop in the extra dimension appears as an adjoint scalar field. Here
we want to obtain pure 4-d Yang-Mills theory (without charged scalars) after 
dimensional reduction. This can be achieved if one does not impose Gauss' law
for the states propagating in the fifth dimension, because the Polyakov loop
is a Lagrange multiplier field that enforces the Gauss law. Formally,
this can be realized simply by putting the fifth component of the gauge 
potential to zero, i.e.
\begin{equation}
A_5 = 0.
\end{equation}
In the infinite volume limit of the 5-d theory this restriction has no effect
on the dynamics. With finite extent in the fifth direction, however, it
deviates from the standard formulation of gauge theories. The leading terms in
the effective action of the 5-d gauge theory take the form
\begin{equation}
S[A_\mu] = \int_0^\beta dx_5 \int d^4x \ \frac{1}{2 e^2}[\mbox{Tr} \ 
F_{\mu\nu} F_{\mu\nu} + \frac{1}{c^2} \mbox{Tr} \ 
\partial_5 A_\mu \partial_5 A_\mu].
\end{equation}
Here $e$ is the dimensionful gauge coupling, which is analogous to $\rho_s$ 
from eq.(\ref{spinaction}) for quantum antiferromagnets, and $c$ is the
velocity of light of the 5-d theory. Note that $\mu$ runs over 4-d indices
only. At finite $\beta$ the above theory has
only a 4-d gauge invariance, because we have fixed $A_5 = 0$, i.e. we have not
imposed Gauss' law. On the other hand, for $\beta = \infty$ a full 5-d gauge
symmetry is recovered, although the above action then still is in $A_5 = 0$ 
gauge. Since we are interested in dimensional reduction, a 4-d gauge symmetry
is sufficient for our purposes. In analogy to eq.(\ref{O3coupling}) for large
$\beta$ the gauge coupling of the induced 4-d theory is given by
\begin{equation}
1/g^2 = \beta/e^2.
\end{equation}

For spin models the Mermin-Wagner-Coleman theorem implies that in the 3-d 
theory with finite extent in the third direction the Goldstone bosons acquire
mass nonperturbatively. In gauge theories, on the other hand, an analogous
theorem, stating that massless gauge bosons cannot exist in four dimensions
unless they do not interact with each other, has not yet been proven. In fact, 
proving such a theorem would mean proving confinement. However, we can turn the
argument around and use the confinement hypothesis to argue that the 
dimensionally reduced 5-d theory indeed has a nonperturbatively generated mass
gap. Using the $\beta$-function of $SU(2)$ gauge theory, in analogy with the 
$O(3)$-model, we expect
\begin{equation}
m \propto \exp(- \frac{12 \pi^2 \beta}{11 e^2}).
\end{equation}
Thus starting from a 5-d nonabelian gauge theory in the massless phase one can
obtain the corresponding 4-d nonabelian gauge theory by making the extent 
$\beta$ of the fifth dimension finite. In fact, $\beta$ plays the role of the
inverse gauge coupling of the 4-d theory, and hence, due to asymptotic 
freedom, we are interested in the large $\beta$ limit. As before, in contrast
to naive expectations, dimensional reduction occurs when the extent of the
fifth dimension becomes large. Again, we want to emphasize that it was 
important not to impose Gauss' law, i.e. to put $A_5=0$. 

Let us also discuss dimensional reduction for Abelian theories. The 2-d
quantum $XY$-model has a Kosterlitz-Thouless transition with a massless phase
at low temperatures. The corresponding 3-d classical model with a finite 
extent in the third direction has an $O(2)$ symmetry with free massless 
particles. In particular, no mass gap is generated in this case. This is not
in conflict with the Mermin-Wagner-Coleman theorem because the particles do
not interact. Again, dimensional reduction occurs --- now already at finite
$\beta$ --- and the resulting 2-d theory is the classical $XY$-model. This is
analogous to what happens in Abelian gauge theories between five and four 
dimensions. A 5-d Abelian gauge theory with finite extent in the fifth 
direction and with $A_5=0$ has massless photons. After dimensional reduction
we end up in the Coulomb phase of 4-d Abelian gauge theory. On the lattice
compact $U(1)$ gauge theory has a phase transition that separates the Coulomb
phase at weak coupling from a confined phase with condensed monopoles. In the
Coulomb phase the monopoles have a mass of the order of the cut-off. Hence, if
one takes the continuum limit anywhere in the Coulomb phase one obtains a
free theory of photons.

Why have we formulated 4-d gauge theory in a 5-d context? First of all, the 
reformulation may shed some light on the structure of fixed points in four and 
five dimensional gauge theories. In the context of standard approaches to 
lattice gauge theory one would probably prefer to work directly in 
four dimensions. However, for quantum link models the situation is different. 
We have defined these models as Hamiltonian theories. Of course, one can also
define a Hamiltonian for ordinary gauge theories. In that case the spectrum
of the Hamilton operator defined on the 3-d space reflects Poincar\'e 
invariance of the corresponding 4-d action. On the other hand, by construction
quantum link models are nonrelativistic gauge theories with no symmetry between
space and time. Hence, a quantum link model with a Hamilton operator defined on
a 3-d spatial lattice can in general not describe a system of elementary 
particles, simply because its spectrum does not reflect Poincar\'e invariance. 
However, in analogy to 2-d quantum spin systems, we expect that in the special
case of quantum link models defined on a 4-d lattice, there are massless modes 
with a relativistic dispersion relation characterized by the velocity of light 
$c$ of the corresponding 5-d theory. The massless 
modes correspond to the deconfined gauge bosons in the weak coupling phase of a
5-d gauge theory. Of course, we cannot interpret the fifth Euclidean direction 
as time, and hence the spectrum of the 4-d Hamilton operator of the quantum 
link model is not the physical spectrum. However, when we make the extent 
$\beta$ of the extra dimension finite, we can make use of the above scenario of
dimensional reduction, in which $\beta$ plays the role of the inverse coupling 
constant of the resulting 4-d gauge theory. In that case one must not impose 
Gauss' law ($\vec G_x |\Psi \rangle = 0$), i.e. gauge variant states also 
propagate in the fifth direction. We are then interested in the quantum
statistical partition function
\begin{equation}
\label{gaugeZ}
Z = \mbox{Tr} \exp(- \beta H).
\end{equation}
In contrast to the standard formulation of gauge theories we have not included
a projection operator on gauge invariant states. From this point of view the 
Hamilton operator of the quantum link model is defined
on a 4-d space-time lattice, and describes the evolution of the system in the
fifth unphysical direction. In particular, all the information about the
physical spectrum of the 4-d theory is contained in correlation functions in 
the Euclidean time direction, which is part of the 4-d lattice. In the
continuum limit $g \rightarrow 0$, which we approach by increasing the extent
$\beta$ of the fifth dimension, we are probing the low lying states in the
spectrum of the 4-d Hamilton operator of the quantum link model. The space-time
correlations in these unphysical states of the 4-d Hamiltonian contain the 
information about the physical spectrum. 

The partition function of 
eq.(\ref{gaugeZ}) can be written as a 5-d path integral of discrete variables
--- in the $SU(2)$ case the eigenstates of the diagonal generators of $SO(5)$
on each link. In many respects this path integral resembles that of quantum
spin systems, which can be simulated by very efficient loop cluster algorithms.
Due to the discrete nature of the Hilbert space, one can even work directly in
the continuum for the extra Euclidean direction \cite{Bea96}. It is plausible
that cluster algorithms can also be constructed for quantum link models, 
which would allow high precision simulations in gauge theories.

\section{Conclusions}

Quantum link models are another class of lattice gauge theories with 
applications
in particle and possibly also in condensed matter physics. In the context
of particle physics quantum link models formulated with a fifth Euclidean
dimension of finite extent resemble ordinary 4-d gauge theories. 
From the point of view of numerical simulations it may seem easier to work
directly in four dimensions using the standard formulation of lattice gauge
theories. However, it could be advantageous to work in five dimensions, 
because the existence of cluster algorithms seems plausible for quantum link 
models. Due to the discrete nature of quantum link models, a discretization of 
the fifth direction is not even necessary. The path integral can be defined 
directly in the continuum, and can perhaps be simulated with an algorithm 
analogous to the one for quantum spins \cite{Bea96}.

Although in this paper we have presented explicit constructions only for pure
$U(1)$ and $SU(2)$ quantum link models, it is straightforward to construct 
models for other gauge groups, and with couplings to charged matter fields. In 
fact, we have also constructed $U(N)$ quantum link models, quantum Higgs 
models, as well as quantum $CP(N)$-models \cite{Cha96}. The inclusion of 
quarks is a nontrivial issue, which is presently under investigation.

At present, we can only speculate about applications of quantum link models to
condensed matter physics. However, due to their general structure, we believe
that they will be at least as useful as ordinary gauge theories. As we have
seen, there are close analogies between quantum spin systems
in two dimensions and quantum link models in four dimensions.
Perhaps there are similar analogies between 1-d quantum spin chains and 
3-d quantum link models. In fact, we find it plausible that Haldane's
conjecture has a gauge analog. Perhaps a 3-d $SU(2)$ quantum link model with 
the spinorial representation of $SO(5)$ on each link corresponds to a 4-d
$SU(2)$ lattice gauge theory with a nontrivial $\theta$-vacuum angle, while in
the vector representation the corresponding vacuum angle might vanish. If so, 
one could learn about the effect of $\theta$ by solving the 3-d quantum link 
model, which may be possible if cluster algorithms become available. One can
also imagine to extend Haldane's conjecture to 1-d quantum $CP(N)$-models in a
similar way. Furthermore, quantum link models allow us to gauge the standard
quantum models of condensed matter physics, for example the Heisenberg model or
the Hubbard model. Such models equipped with a lattice gauge symmetry may 
eventually be useful to describe phenomena like, for example, high $T_c$ 
superconductivity or the quantum Hall effect.

In conclusion, there is a whole class of models in various dimensions and
with various symmetries that are of theoretical, and in some cases even of 
phenomenological interest. A lot of work needs to be done before it will be
clear how useful quantum link models are in particle and condensed matter
physics.

\section*{Acknowledgements}

We are indebted to W. Bietenholz, R. Brower and J. Goldstone for very 
helpful conversations. We also like to thank P. Orland, who drew our attention 
to refs.\cite{Hor81,Orl90,Orl92}, for very interesting discussions. One of the 
authors (U.-J. W.) likes to thank the theory group of Erlangen University, 
where part of this work was done, for its hospitality, and the A. P. Sloan 
foundation for its support.


\begin{thebibliography}{10}
 
\bibitem{Fro93}
J. Fr\"ohlich and U. Studer, Rev. Mod. Phys. 65 (1993) 733.

\bibitem{Eck89}
U. Eckern and A. Schmid, Phys. Rev. B39 (1989) 6441; \newline
R. Fazio and G. Sch\"on, Phys. Rev. B43 (1991) 5307; \newline
M. C. Diamantini, P. Sodano and C. A. Trugenberger, Nucl. Phys. B474 (1996) 
641.

\bibitem{Hei28}
W. Heisenberg, Z. Phys. 49 (1928) 619.

\bibitem{Hor81}
D. Horn, Phys. Lett. 100B (1981) 149.

\bibitem{Orl90}
P. Orland and D. Rohrlich, Nucl. Phys. B338 (1990) 647.

\bibitem{Orl92}
P. Orland, Nucl. Phys. B372 (1992) 635.

\bibitem{Hal83}
F. D. M. Haldane, Phys. Lett. 93A (1983) 464; Phys. Rev. Lett. 50 (1983) 1153;
J Appl. Phys. 57 (1985) 33.

\bibitem{Bet31}
H. Bethe, Z. Phys. 71 (1931) 205.

\bibitem{Lie61}
E. H. Lieb, T. Schultz and D. J. Mattis, Ann. Phys. 16 (1961) 407; \newline
I. Affleck and E. H. Lieb, Lett. Math. Phys. 12 (1986) 57; \newline
I. Affleck, T. Kennedy, E. H. Lieb and H. Tasaki, Phys. Rev. Lett. 59 (1987)
799; Commun. Math. Phys. 115 (1988) 477.

\bibitem{Bot84}
R. Botet, R. Jullien and M. Kolb, Phys. Rev. B30 (1984) 215; \newline
J. B. Parkinson and J. C. Bonner, Phys. Rev. B32 (1985) 4703; \newline
M. P. Nightingale and H. W. J. Bl\"ote, Phys. Rev. B33 (1986) 659; \newline
U. Schollw\"ock and T. Jolicouer, Europhys. Lett. 30 (1995) 493.

\bibitem{Bie95}
W. Bietenholz, A. Pochinsky and U.-J. Wiese, Phys. Rev. Lett. 75 (1995) 4524.

\bibitem{Nov81}
S. P. Novikov, Sov. Math. Dokl. 24 (1981) 222; Usp. Math. Nauk. 37 (1982) 3;
\newline
E. Witten, Commun. Math. Phys. 92 (1984) 455.

\bibitem{Aff88}
I. Affleck, in Fields, Strings and Critical Phenomena, Proceedings of the
Les Houches Summer School, Session XLIX, edited by E. Brezin and J. Zinn-Justin
(North Holland, Amsterdam, 1988), p. 563.

\bibitem{Bar91}
T. Barnes, Int. J. Mod. Phys. C2 (1991) 659.

\bibitem{Wie94}
U.-J. Wiese and H.-P. Ying, Z. Phys. B93 (1994) 147.

\bibitem{Bea96}
B. B. Beard and U.-J. Wiese, Phys. Rev. Lett. 77 (1996) 5130.

\bibitem{Leu90}
P. Hasenfratz and H. Leutwyler, Nucl. Phys. B343 (1990) 241.

\bibitem{Mer66}
N. D. Mermin and H. Wagner, Phys. Rev. Lett. 17 (1966) 1133; \newline
S. Coleman, Commun. Math. Phys. 31 (1973) 259.

\bibitem{Has90}
P. Hasenfratz, M. Maggiore and F. Niedermayer, Phys. Lett. B245 (1990) 522;
\newline
P. Hasenfratz and F. Niedermayer, Phys. Lett. B245 (1990) 529.

\bibitem{Has91}
P. Hasenfratz and F. Niedermayer, Phys. Lett. B268 (1991) 231.

\bibitem{Cha89}
S. Chakravarty, B. I. Halperin and D. R. Nelson, Phys. Rev. B39 (1989) 2344.

\bibitem{Eve93}
H. G. Evertz, G. Lana and M. Marcu, Phys. Rev. Lett. 70 (1993) 875.

\bibitem{Wol89}
U. Wolff, Phys. Rev. Lett. 62 (1989) 361; Nucl. Phys. B334 (1990) 581.

\bibitem{Din90}
H.-Q. Ding and M. S. Makivi\'c, Phys. Rev. B42 (1990) 6827.

\bibitem{Cre79}
M. Creutz, Phys. Rev. Lett. 43 (1979) 553.

\bibitem{Cha96}
S. Chandrasekharan and U.-J. Wiese, in preparation.

\end{thebibliography}
\end{document}